\documentclass[11pt]{article}
\usepackage{amssymb,amsmath,graphicx,epsfig}
\usepackage{graphicx}
\begin{document}
\newcommand{\M}{\mbox{m}} \newcommand{\n}{\mbox{$n_f$}}
\newcommand{\EL}{\mbox{e}} \newcommand{\EP}{\mbox{e$^+$}}
\newcommand{\EM}{\mbox{e$^-$}} \newcommand{\EPEM}{\mbox{e$^+$e$^{-}$}}
\newcommand{\EMEM}{\mbox{e$^-$e$^-$}}
\newcommand{\GG}{\mbox{$\gamma\gamma$}}
\newcommand{\GE}{\mbox{$\gamma$e}}
\newcommand{\GP}{\mbox{$\gamma$e$^+$}} \newcommand{\TEV}{\mbox{TeV}}
\newcommand{\MEV}{\mbox{MeV}} \newcommand{\GEV}{\mbox{GeV}}
\newcommand{\LGG}{\mbox{$L_{\gamma\gamma}$}}
\newcommand{\LGE}{\mbox{$L_{\gamma e}$}}
\newcommand{\LEE}{\mbox{$L_{ee}$}}
\newcommand{\WGG}{\mbox{$W_{\gamma\gamma}$}}
\newcommand{\EV}{\mbox{eV}} \newcommand{\CM}{\mbox{cm}}
\newcommand{\MM}{\mbox{mm}} \newcommand{\NM}{\mbox{nm}}
\newcommand{\MKM}{\mbox{$\mu$m}} \newcommand{\SEC}{\mbox{s}}
\newcommand{\CMS}{\mbox{cm$^{-2}$s$^{-1}$}}
\newcommand{\MRAD}{\mbox{mrad}}
\newcommand{\IND}{\hspace*{\parindent}}
\newcommand{\E}{\mbox{$\varepsilon$}}
\newcommand{\EN}{\mbox{$\varepsilon_n$}}
\newcommand{\EI}{\mbox{$\varepsilon_i$}}
\newcommand{\ENI}{\mbox{$\varepsilon_{ni}$}}
\newcommand{\ENX}{\mbox{$\varepsilon_{nx}$}}
\newcommand{\ENY}{\mbox{$\varepsilon_{ny}$}}
\newcommand{\EX}{\mbox{$\varepsilon_x$}}
\newcommand{\EY}{\mbox{$\varepsilon_y$}}
\newcommand{\BI}{\mbox{$\beta_i$}} \newcommand{\BX}{\mbox{$\beta_x$}}
\newcommand{\BY}{\mbox{$\beta_y$}} \newcommand{\SX}{\mbox{$\sigma_x$}}
\newcommand{\SY}{\mbox{$\sigma_y$}}
\newcommand{\SZ}{\mbox{$\sigma_z$}}
\newcommand{\SI}{\mbox{$\sigma_i$}}
\newcommand{\SIP}{\mbox{$\sigma_i^{\prime}$}}
\newcommand{\MW}{\mbox{MW}} \newcommand{\CE}{\mbox{$\mathcal{E}$}}
\title{Critical Issues in Linear Colliders~\thanks{Talk at 
Workshop on Quantum Aspects of Beam Physics and Other Critical
Issues of Beams in Physics and Astrophysics, January 7--11, 2003,
Hiroshima University, Higashi-Hiroshima, Japan}}

\author{Valery Telnov  \\
{\it Institute of Nuclear Physics, 630090 Novosibirsk, Russia}}
    
\date{} 
\maketitle
\begin{abstract} 
Linear colliders (LC) on the energy 0.5--1 TeV are
  considered as the next step in the particle physics. High
  acceleration gradients, small beam sizes, precision tolerances, beam
  collision effects are main problems for linear colliders.  In this
  paper we discuss physics motivation, parameters and status of
  current LC projects, \EPEM, \GG\ and \GE\ modes of operation,
  physical limitations on the energy and luminosity.  Present
  technologies allow to reach energies about 5 TeV with adequate
  luminosities.  Advanced technique based on plasma and laser method
  of acceleration can provide much higher accelerating gradients,
  however, perspectives of these methods for high energy colliders are
  still under big question.  Linear colliders with energies above 10
  TeV are hard for any acceleration technology. Speculations on
  possibility of PeV linear colliders based on ponderomotive laser
  acceleration are just not serious and contain several mistakes on
  conceptual level.  It is shown that due to radiation in the
  transverse laser field,  methods of acceleration based on laser bunch
  ``pressure'' do not work at  high energies. 
\end{abstract}

\section{Introduction: next steps in particle physics}

Progress in particles physics in the last several decades was
connected with the increase of accelerator energies.  Historically,
two types of colliders co-existed and gave main results,
$pp(p\bar{p})$ and \EPEM. Proton colliders give access to higher
energies, but \EPEM\ colliders have simple initial state, smaller
background and allow much better precision. At proton colliders
$c,b,t$ quarks and $W,Z$ bosons have been discovered, while at \EPEM\ 
colliders $c$-quark, $\tau$-lepton, gluon. In addition, at \EPEM\ colliders
 $c,b,W,Z,\tau$ physics has been studied  with a high
accuracy providing a precision test of the Standard Model.

The next proton collider LHC with the energy $2E_0=14$ TeV will start
operation in about 2007. It will certainly bring new discoveries. But,
as before, for detail study of new physics and it's {\it
  understanding} a \EPEM\ collider is very desirable. Such projects on
the energy $2E_0=$0.5--1.5 TeV already exist, but, unfortunately,
approval is delayed due to a high cost and necessity of 
international cooperation. According to present understanding the
construction can start in about 2007.

As for long-term perspectives of particle physics, the future is even less
clear. Three kind of facilities are under discussion: Very Large
Hadronic Collider (VLHC) with $pp$ beams on the energy up to 200
TeV, Compact \EPEM\ Linear Collider CLIC on the energy $2E_0 =$ 3--5 TeV
and muon colliders which potentially can reach a c.m.s. energy even
higher than in $pp$ collisions. 
  
Physics motivation for next generation of colliders (LHC, LC) is very
strong, two  examples are given below.

If the Standard Model is valid a new particle, the Higgs boson, should
exist. Direct search at LEP and measurements of loop corrections
indicate that the Higgs boson mass lays in the region 115--200 GeV.
Such a particle should have very special properties, their coupling
constants with other particles are proportional to particle masses.
Linear colliders allow us to measure Higgs branchings with a high
accuracy, So, experiments at LHC and LC can shed a light on the origin
of particle masses.

The second physics goal is a search of a supersymmetry which assumes
the existence of a new class of particles, superpartners of known
particles but with different spins: particles with the spin 1/2 have
partners with the spin 0 and vice versa.  It is possible that the dark
matter in the universe consists of the lightest neutral supersymetrical
particles. At colliders, one could produce any kind of such particles,
charged and neutral. A discovery of a ``parallel'' world (which
according to astronomical data has a density even higher than that of
the barionic matter) would mean a new revolution in physics.

Below we consider existing projects of linear colliders, their
problems, energy and luminosity limitations, prospects of advanced
accelerator methods.

\section{Projects of linear colliders} \label{projects}

It was realized already 30 years ago that the energy of circular
\EPEM\ linear colliders is limited by synchrotron radiation losses at
a level of 100--200 GeV and further progress is only possible using
linear \EPEM\ colliders~\cite{skrinsky}.  At the end of 1980-th the
2-mile electron linac at SLAC has been transformed into a (semi)linear
collider SLC with the c.m.s.  energy of 90 GeV. It gave nice physics
results and a great experience of work at the first linear collider.

\begin{figure}[p]
\begin{center} \vspace*{-0cm}
\includegraphics[width=5.2cm,bb= 118 114 414 654,clip,angle=-90]{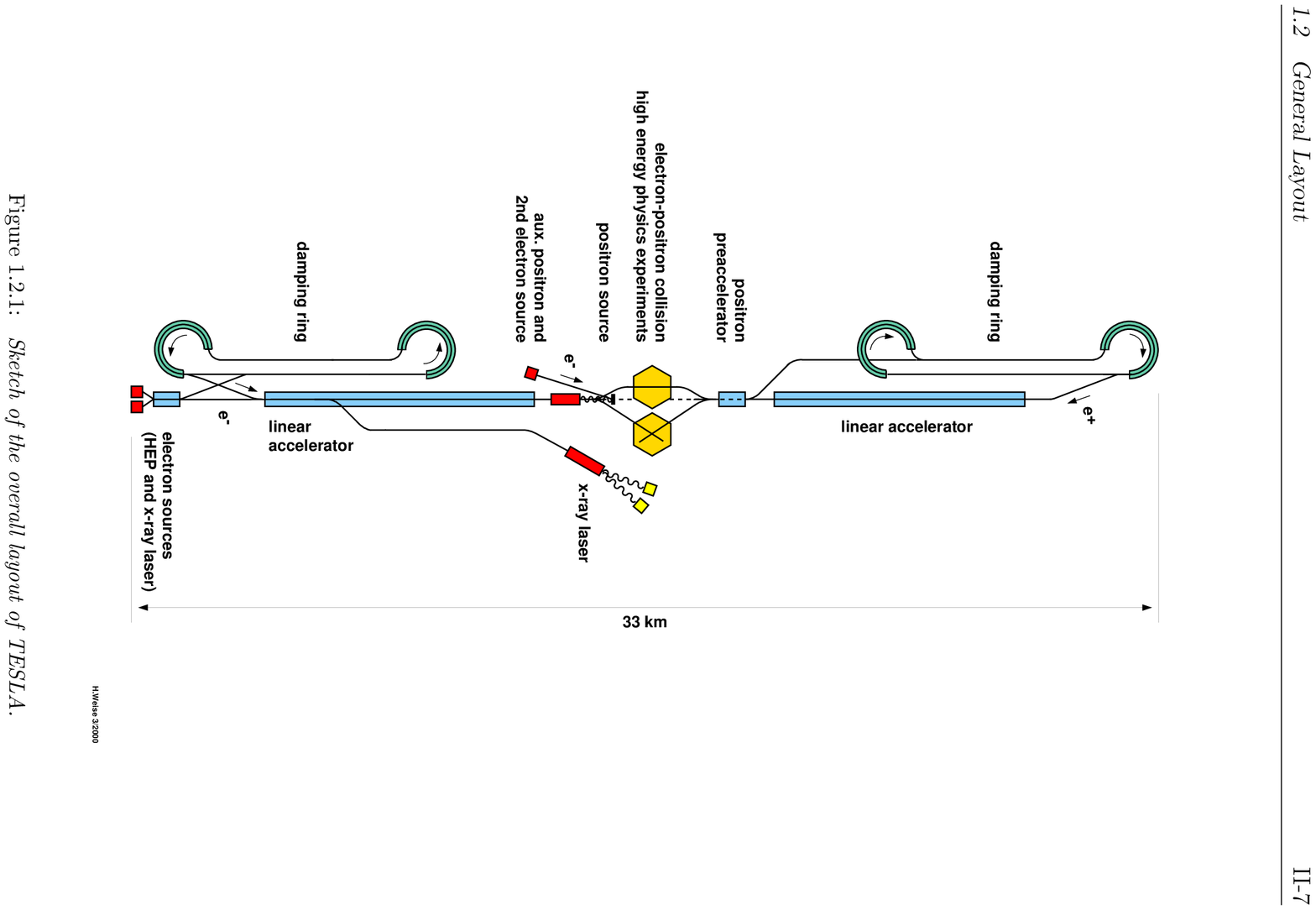} \\
\vspace*{-2cm}\includegraphics[width=7.2cm,bb= 66 194 473 673,clip,angle=-90]{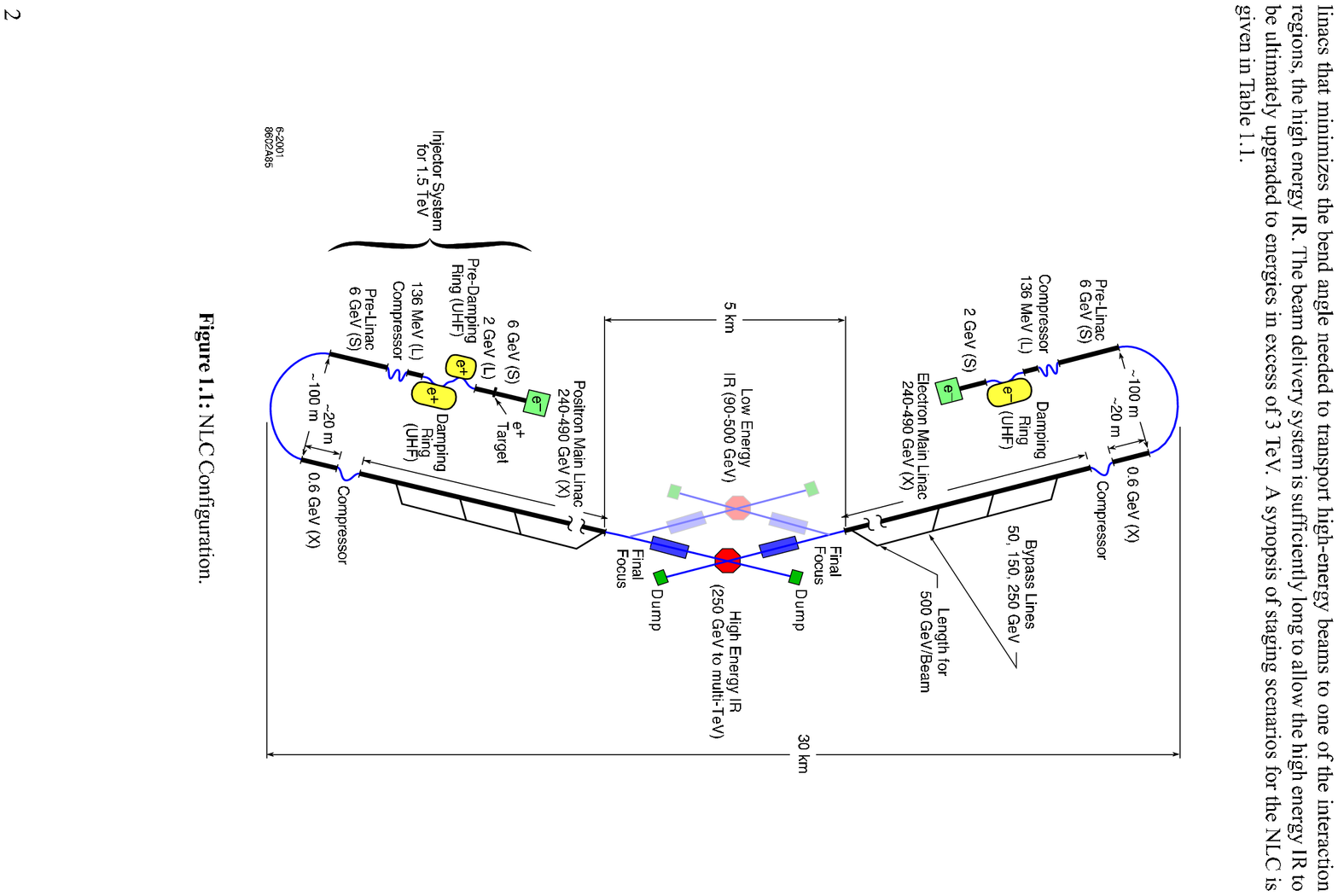} \\
\vspace*{-1cm}\includegraphics[width=4.7cm,bb= 166 50 535 777,clip,angle=-90]{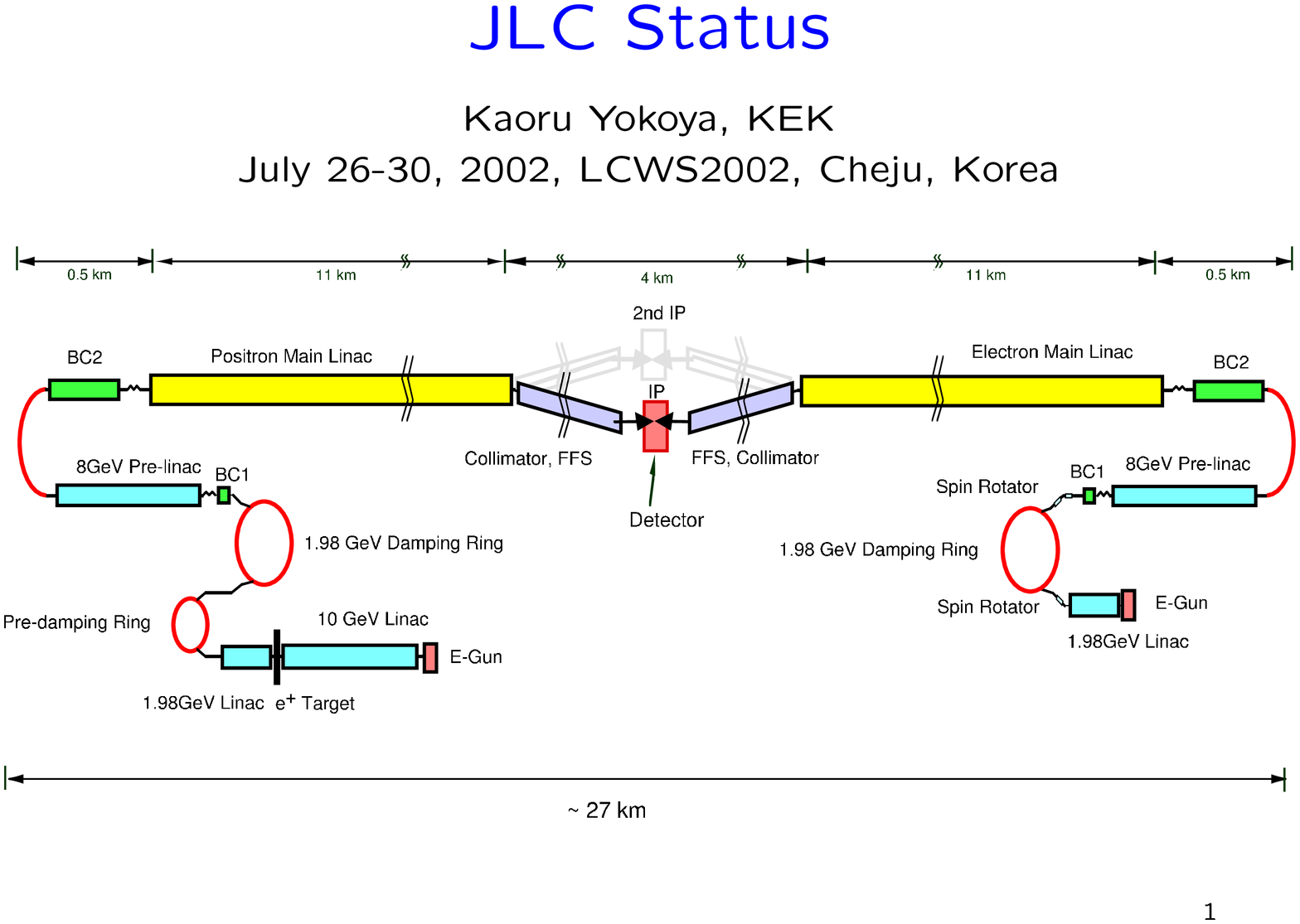} \\
\vspace*{-1cm}\includegraphics[width=7.2cm,bb= 66 194 473
673,clip,angle=-90]{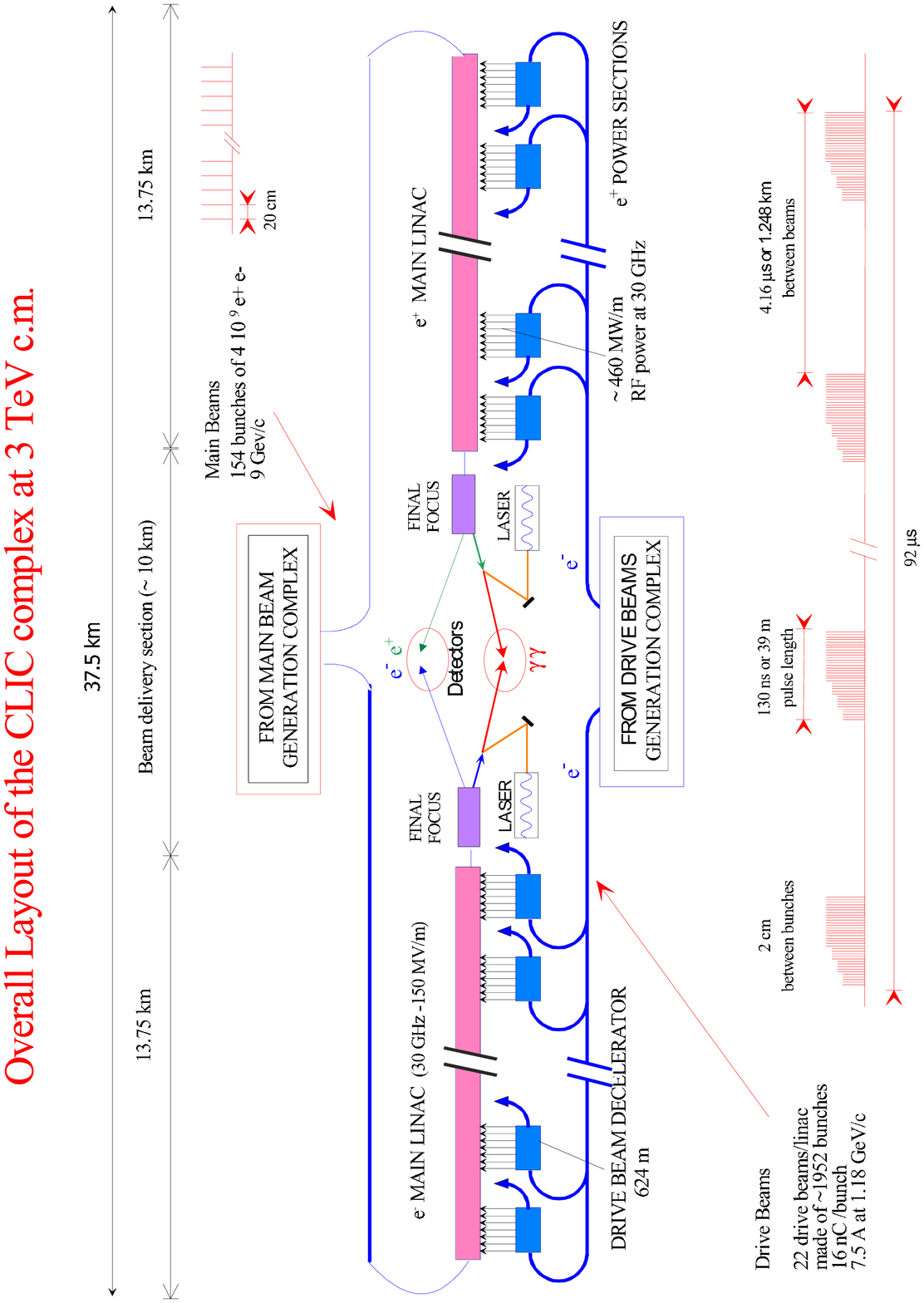}  \\
\end{center}
\caption{Schemes of linear colliders TESLA, NLC, JLC and CLIC (from up
  to down).}
\label{colliders}
\end{figure}

At the same time an international study on linear collider lead by
SLAC, KEK, DESY, CERN and BINP has been launched with ambitious goal to
develop a linear collider with an energy about one TeV and a
luminosity by a factor of $10^{3}$--$10^{4}$ higher than it was at the SLC.
Since that time a lot of developments have been done and now three
projects TESLA~(Europe)~\cite{TESLA}, NLC~(US)~\cite{NLC},
JLC~(Japan)~\cite{JLC} are almost ready for construction.
A fourth project CLIC(CERN)~\cite{CLIC} is focused on multi-TeV energies
and is considered as the next-to-next linear collider.
Schemes of colliders are shown in Fig.~\ref{colliders}, main
parameters are presented in Table~\ref{parameters}.
\begin{center}

\begin{table}[!htb]
\caption{Parameters of linear collider}
{\renewcommand{\arraystretch}{1.0} \setlength{\tabcolsep}{1mm}
 \small 

\vspace{0.5cm}

\hspace*{0.7cm}
\begin{tabular}{l | l| c c | c c | c c } 
 \multicolumn{2}{c}{}   &\multicolumn{2}{c}{{\normalsize  TESLA}}  &
 \multicolumn{2}{c}{{\normalsize JLC/NLC}} &
 \multicolumn{2}{c}{{\normalsize CLIC}} \\ 
&&&&&&& \\
2E$_0$    & GeV & 500 & 800 & 500 & 1000 & 500 & 3000 \\ \hline \hline
Site L & km& --  & 33 & -- & 32 & -- & 40 \\ 
Two linac L & km& 30 & 30 & 12.6 & 25.8 & 5 & 27.5 \\ 
Beam del. L & km& 3.2 & 3.2 & 3.8 & 3.8 & 5 & 5  \\ 
G(un.l/load) &MeV/m& 23.4 & 35 & 70/55 & 70/55 & 172/150 & 172/150 \\
Total AC  &MW& 95 & 160 & 120 & 240 & 100 & 300 \\
AC-beam eff. &\%& 23 & 21 & 10 & 10 & 8.5 & 8.5 \\
RF freq. & GHz& 1.3 & 1.3 & 11.4 & 11.4 & 30 & 30 \\ \hline
Rep. rate & Hz& 5 & 4 & 120 & 120 & 200 & 100 \\
bunch/train &  & 2820 & 4886 & 192 & 192 & 154 & 154 \\
Coll. rate & kHz & 14.1 & 19.5 & 23 & 23 & 30.8 & 15.4 \\
Bunch separ. & ns & 337 & 176 & 1.4 & 1.4 & 0.67 & 0.67 \\ 
Train length & $\mu$sec & 950 & 860 & 0.267 & 0.267 & 0.1 & 0.1 \\ \hline 
Part./bunch & $10^{10}$ & 2 & 1.4 & 0.75 & 0.75 & 0.4 & 0.4 \\  
$\sigma_z$  & $\mu$m   & 300 & 300 & 110 & 110 & 30 & 30 \\ 
\ENX/\ENY\ & mm$\cdot$mrad & 10/0.03  & 8/0.015   & 3.6/0.04  & 3.6/0.04  & 2/0.02    & 0.68/0.02 \\ 
\BX/\BY\  & mm   & 15/0.4 & 15/0.4 & 8/0.11 & 13/0.11 & 10/0.15 & 8/0.15 \\
$\sigma_x/\sigma_y$  & nm   & 553/5 & 391/2.8 & 243/3 & 219/2.3 &
200/2.5 & 43/1 \\ \hline
$D_x/D_y$  &    & 0.2/25 & 0.2/27 & 0.16/12.9 & 0.08/10 & 0.12/7.9
&0.03/2.7 \\ 
$\Upsilon_0$  &    & 0.06 & 0.09 & 0.14 & 0.29 & 0.3 & 8.1 \\ 
$\delta$  & \%   & 3.2 & 4.3 & 4.7 & 8.9 & 3.8 & 31 \\
$n_{\gamma}$/e  &   & 2 & 1.5 & 1.3 & 1.3 & 0.7 & 2.3 \\
$n_{\EPEM}$/e  &   &  &  &  &  &  & 0.17 \\ \hline
L(with pin.)  &$10^{34}$\,\CMS &3.4  &5.8  & 2  & 3  & 1.4 & 10.3 \\ 
L(w/o pin.)  &$10^{34}$\,\CMS &1.6  &2.8  &1.2  &1.9  & ? & ? \\ 
L(1\%)/L& \% & 66 &   & 64  &   & 67 & 25.5 \\
L(5\%)/L& \% & 91 &   & 85  &   & 86 & 40.8 \\ 
\end{tabular}}
\label{parameters}
\end{table}
\end{center}
\vspace{-5mm}
Each project has some distinctive features:\\[2mm]
$\bullet$ TESLA: L band, 1.4 GHz, superconducting, $G_{max} \sim 35$
MeV/m, a good efficiency, a low wakefield, a relaxed alignment
tolerances,
a large distance between bunches; \\
$\bullet$ NLC/JLC: X-band, 11.4. GHz, warm cavities, a high gradient
(55
MeV/m loaded); \\
$\bullet$ CLIC: 30 GHz, a two-beam accelerator (one of beams produces
RF power), a very high gradient, 150 MeV/m, cost effective at
multi-TeV energies.

  So, there are three main technologies for LC developed by large
  teams, each project have certain advantages. It would be good
  to built two colliders almost simultaneously: TESLA for  energies
  below 0.5 TeV, NLC/JLC for the energy region up to 1.5 TeV and a
  third collider, CLIC, on the energy 3--5 TeV one decade later. 
  However, due to a high cost only one global linear collider is seen
  in the visible future. 

\section{General features of linear colliders}

At storage rings, each bunch collides many times, the RF power is
spent mainly for compensation of synchrotron radiation losses. At
linear colliders, each bunch is used only once, radiation losses
during the acceleration are negligible, but a lot of energy is needed
for production and acceleration of bunches with a high rate. The total
RF power consumption at LEP and at 0.5 TeV linear colliders are
comparable, of the order of 100 MW from the wall plug.

The number of accelerated particles is limited
by total AC power which is proportional to the beam power $P$. Due
to the dependence of cross sections on the energy as $\sigma \propto
1/E^2$ the luminosity should increase as $E^2$, as a result the required
transverse beam sizes  at TeV energies should be very small.
 
Beams with small sizes have very strong fields that lead to large
radiation losses during beam collisions (beamstrahlung).  This effect
does not allow us to use beams with simultaneously small horizontal and
vertical beam sizes ($\sigma_x, \sigma_y$) (only very flat beams) and
to get the required luminosity the beam power should be additionally
increased.  This leads to the ``energy crisis'' at the beam energy of
about $2E_0 \sim 5\, \TEV$, see Sec.~4. In the \GG\ mode of operation
(Sec.~5) only somewhat higher energies are possible due to conversion
of high energy photons to \EPEM\ pairs in the field of the opposing
beam (coherent pair creation).

Beside traditional linear accelerators, there are ideas of using plasma and
laser high gradient accelerator techniques for linear colliders. 
There are some speculations about colliders with 100 TeV
and even PeV energies. Certainly, development of these techniques will
lead to some practical applications, but obtaining  colliding beams
is very problematic due to required  quality of beams and collision
effects. Some considerations and critical remarks on plasma and laser
acceleration are given Sec.~\ref{advanced}.

\section{Collision effects restricting luminosity and energy of
  linear colliders} \label{collision}
  
In order to obtain a sufficient luminosity at linear colliders the
beam sizes should be very small. This causes two sorts of problems: a)
generation and acceleration of beams with very small emittances and
focusing to a tiny spot, b) beam-beam collision effects which lead to
degradation of the beam quality.

The first problem is very difficult but not fundamental, in principle,
one can obtain emittance smaller than give damping rings using, for
example, laser cooling.  The second problem is even more
severe: beam collision effects put restrictions on attainable
luminosity and, correspondently, on the maximum energy of linear
colliders.
    
  In the absence of collision effects the luminosity of a collider
\begin{equation}
L \approx  \frac{N^2 f}{4\pi\sigma_x\sigma_y} = \frac{P}{4\pi
  E_0}\; \times \;\frac{N}{\sigma_x \sigma_y}\,.
\label{lumi}
\end{equation}
For $2P=20$ \MW\ (200 \MW\ AC power), $N=2\times 10^{10},
\;\sigma_x=\sigma_y= 1$ nm it gives $L= 10^{37}/E_0[\TEV],\;\CMS,$
this luminosity is sufficient for production of $10^3$ lepton pairs per $10^7$
sec up to  $2E_0 = 25$ TeV. Below we consider several 
limitations due to collisions effects.

\subsection{Pinch effect and instability of beam collisions}

During the collision beams attract (\EPEM) or repulse (\EMEM) each
other. The characteristic disruption parameter~\cite{solyak,Yokoya}
\begin{equation}
 D_y = \frac{2Nr_e\sigma_z}{\gamma \sigma_x \sigma_y}\,.
\end{equation}
For flat beam and $D_y \sim 10$, the attraction  leads to
increase of the \EPEM\ luminosity by a factor of $H_D \sim 2$. At $D_y \ge
25$ beams become unstable, the corresponding luminosity 
\begin{equation}
L \sim \frac{P}{mc^2 r_e \sigma_z}\,.
\end{equation}
For $P= 10$ MW and $\sigma_z = 100\;\MKM$ $L \sim 5\times 10^{34}\;\CMS$.
So,  this put limit on the luminosity for a given beam power and bunch
length.

\subsection{Beamstrahlung}
A strength of a beam field is characterized by the parameter
$\Upsilon$~\cite{Noble,Yokoya}
\begin{equation}
\Upsilon = \frac{2}{3}\frac{\hbar \omega_c}{E} = \gamma
\frac{B}{B_0}, \;\;\;\; B_0 = \frac{\alpha e}{r_e^2} = 4.4\times
10^{13} \mbox{G}\,.
\end{equation}
For flat beams
\begin{equation}
\Upsilon_{av} \approx \frac{5 N r_e^2 \gamma}{6\alpha
  \sigma_x \sigma_z}\,.
\end{equation}
The maximum value of $\sigma_z$ is determined by disruption. Ideally,
increasing $\sigma_x$ to infinity and simultaneously decreasing
$\sigma_y$ to zero one can get arbitrary small $\Upsilon$ for any
luminosity. However, if $\sigma_y$ has some minimum value (there are
many reasons), then $\Upsilon \propto L^2 \gamma^2 \sigma_y/P^2 D_y$.
As $P$ is always limited, $D_y<25$ and the required $L$ increases
with the energy as $\gamma^2$, the  value of $\Upsilon$ increases rapidly
with the energy. In the current LC projects at  1 TeV
$\Upsilon =  {\mathcal{O}}(1)$, at higher energies inevitably
 $\Upsilon \gg 1$. 
 
 Synchrotron radiation of electrons in the field of the opposing beam
 (beamstrahlung) put severe limitations on performance of linear
 colliders. Energy losses are given by  approximate
 formulae~\cite{Yokoya}: 
\vspace{-0mm}
\begin{equation}  \frac{dN_{\gamma}}{dt} = \frac{5}{2\sqrt{3}}\frac{\alpha^2 c
    \Upsilon}{r_e \gamma} U_0(\Upsilon),\;\;\;\;\;\;U_0 \approx
  \frac{1}{(1+\Upsilon^{2/3})^{1/2}},
\end{equation} \vspace{-0mm}
\begin{equation} 
-\frac{dE}{E dt} = \frac{2}{3}\frac{\alpha^2 c \Upsilon^2}{r_e
  \gamma} U_1(\Upsilon) \;\;\;\;\; U_1 \approx
  \frac{1}{(1+(1.5\Upsilon)^{2/3})^2},
\end{equation} \vspace{-0mm}
\begin{equation} 
\frac{<\omega>}{E} = \frac{4\sqrt{3}}{15} \Upsilon
\frac{U_1(\Upsilon)}{U_0(\Upsilon)} \;\;= 0.462 \Upsilon 
\;\;(\Upsilon \to 0),\;\;\;\;
0.254 \;\; (\Upsilon \to \infty),
\end{equation} \vspace{-0mm}
\begin{equation} 
\delta_E=\frac{\Delta E}{E} = 1.24\left[\frac{\alpha^2 \sigma_z
    \Upsilon}{r_e \gamma} \right]\Upsilon U_1(\Upsilon);
\label{delta}
\end{equation} 
$\Upsilon \ll 1$ is the ``classic'' regime; $\Upsilon \sim$ 0.2--200 
the ``transition'' regime ($\Upsilon U_1(\Upsilon) \approx$ 0.1--0.2 $\sim
0.15$); $\Upsilon \gg 200 $ the ``quantum'' regime. Colliders in the TeV region
belong to the transition regime, multi-TeV LC with dense short bunches can
reach the quantum regime.

The luminosity (\ref{lumi}) can be expressed via $\delta_E$.
In the transition regime it does not depend on the bunch length $\sigma_z$:
\begin{equation}
L \sim \frac{6.45 \delta_E}{ 4 \pi \alpha r_e \gamma \sigma_y}
\left( \frac{P}{mc^2} \right) = 1.5\times 10^{34}\frac{P[\MW]
 \delta_E}{E_0[\TEV] \sigma_y[\NM]} \;\CMS;
\end{equation}
In the quantum regime 
\begin{equation}
 L \sim \frac{1.95}{4\pi \alpha^2 \sigma_y} \sqrt{\frac{\delta_E^3}{r_e
    \sigma_z \gamma}} \left( \frac{P}{mc^2} \right) = 5 \times 10^{34}
\frac{P[\MW]}{\sigma_y[\NM]}\sqrt{\frac{\delta_E^3}
  {E_0[\TEV]\sigma_z[\MKM]}}.
\end{equation}
For example, for $P=10$ \MW\ per beam (about 200 MW from wall plug)
$\sigma_y=1$ nm, $2E_0=5$ TeV, $\delta_E = 0.2$ we get (accuracy is
about factor of 2--3) $ L= 1.2 \times 10^{34}$ \CMS\ in the transition
regime (does not depend on $\sigma_z$) and $ L= 3 \times 10^{34}$
\CMS\ in the quantum regime (for $\sigma_z=1\;\MKM$), an additional
factor of $\sim 1.5$ can give the pinch effect.  We see that the quantum
regime (short bunches) helps but not too much.

In order to produce $10^3$ characteristic reactions $\EPEM\to
\mu^+\mu^-$ per $10^7$ sec at the energy $2E_0 = 5$ TeV the
required luminosity is $3\times 10^{34}$, that is close to the above
limit due to beamstrahlung.  So, if $\sigma_{y, min} \sim 1$ nm (see
Sec.~\ref{symin}), the maximum reasonable energy of linear
colliders is about $2E_0 \sim 5$ TeV.

In principle, there is a possibility to cancel beam fields by
colliding four beams (\EPEM\ from each side), then beamstrahlung is
absent. The beams instability threshold remains at the same level of
luminosity or may be only somewhat higher. This scheme can give some
gain in luminosity, but technically it looks unrealistic.
  
\subsection{Coherent \EPEM\ pair creation}
At $\kappa = (\omega/E_0)\Upsilon > 1$ a beamstrahlung photon can
convert into \EPEM\ pairs in the field of the opposing
beam~\cite{ChenTel}. At $\kappa \gg\ 1$ the ratio of beamstrahlung/pair
creation probabilities is about 3.8. The number of beamstrahlung
photons at linear colliders $N_{\gamma} \sim N_{\EL}$ (in order to increse
luminosity the horizontal size is decreased until each electron emit
about one photon).  Therefore the number of \EPEM\ pairs at $\kappa
\gg 1$ (or $\Upsilon \gg 1$), $N_{\EPEM}/N_{\EL} = {\mathcal{O}}(0.1).$
For example, at CLIC(3000) $N_{\EPEM}/N_{\EL} \sim 0.085$.  The minimum
energy of produce particles (important from a background point of
view) $E_{min} \sim 0.05 E_0/\Upsilon$.

\subsection{Deflection of soft particles}

The lowest energy charged particles produced in process of coherent
pair creation with the same sign of the charge as that of the
opposing beam are deflected by the opposing beam on the angle~\cite{ChenTel}
\begin{equation}
 \theta \sim \left(\frac{4\pi N e^2}{\sigma_z E_{min}}\right )^{1/2}
\sim 170 \frac{N}{\sigma_z}\left(\frac{r_e^3}{\sigma_x}\right )^{1/2}.
\end{equation}  
For example, at CLIC $\theta \sim 15$ mrad.  To avoid background from
these large angle particles one should use the crab-crossing
scheme~\cite{Palmer}.  Below we will see that crab-crossing angles below
20--30 mrad are acceptable, but larger angles lead to the increase of
the vertical beam size.

So, deflection of soft particles put an additional constraint on the beam
parameters. Beamstrahlung and instabilities may be OK (in case of very
short bunches), but disruption angles are too large.

\subsection{Minimum value of $\sigma_y$} \label{symin}
The minimum vertical beam size at the interaction point (at $\beta_y
\sim \sigma_z$) $\sigma_y= \sqrt{\ENY \sigma_z/\gamma}$.
Limitations:
\begin{itemize}
\item Attainable value of the normalized vertical emittance from an
injector; 
\item Radiation in final quadrupoles (Oide effect)~\cite{oide}. Minimum
achievable beam size $\sigma_{min}[\mbox{m}] \approx 1.7\times
10^{-4}\ENY[\mbox{m}]^{5/7}.$ For \ENY\ considered in the current LC
projects $ \sigma_{min} \sim 0.5$ nm;
\item  Radiation in the detector solenoid field due to the crab
crossing~\cite{NLC_Z0,telnov_s3,zimmer}
\end{itemize}
\begin{equation}
  \sigma_y^2 = \frac{55 r_e^2}{480\sqrt{3}\, \alpha} 
\left( \frac{eB_s\theta_c L}{2 mc^2} \right)^5.
\label{solen}
\end{equation}
For $B_s=4$ T, $L=4$ m \
 $\sigma_{y} = 0.74$ nm for $\theta_c = 20$ mrad and
 2 nm for $\theta_c = 30$ mrad.  
More accurate simulation of this effect  (the number of emitted photon
is about one) was done in Refs~\cite{telnov_s3,zimmer}.
  As a linear collider without a detector has no sense this effect put
  a limit on a minimum vertical beam size at the interaction point at
  the level of 0.5 nm at $\theta_c = 20$ mrad. 

\subsection{Resume on maximum energies of linear colliders}

For a reasonable wall plug AC power 100--300 MW the maximum energy of
linear \EPEM\ colliders with a luminosity sufficient for experiments,
according to present understanding, is limited by collision effects at
the level of $2E_0 =$ 5--10 TeV. 

\section{Photon colliders}

In addition to \EPEM\ physics, linear colliders provide a unique
opportunity to study \GG\ and \GE\ interactions at high energies and
luminosities~\cite{GKST83,GKST84}. High energy photons can be obtained
using Compton backscattering of laser light off high energy electrons.
This option is foreseen in all other project of linear
colliders~\cite{TESLA,NLC,JLC,CLIC,TESLATDR}.
The maximum energy of photons after Compton scattering
\begin{equation}
\omega_m=\frac{x}{x+1}E_0; \;\;\;\; x \approx
\frac{4E_0\omega_0}{m^2c^4} \simeq
15.3\left[\frac{E_0}{\TEV}\right]\left[\frac{\omega_0}{\EV}\right].
\end{equation}
For example: $E_0 =250$~GeV, $\omega_0 =1.17$~eV ($\lambda=1.06$ \MKM)
$\Rightarrow$ $x=4.5$ and $\omega_m = 0.82E_0 = 205$ GeV.  The value
$x = 4.8$ is the threshold for the process $\gamma \gamma_L \to \EPEM$
in the conversion region. This determine the optimum laser wavelength:
$\lambda_{opt} \sim 4 E_0[\TEV]\;\MKM$~\cite{TEL90}. Nonlinear effects in
Compton scattering increase the threshold value of $x$ by a factor of
$(1+\xi^2)$, where a parameter of nonlinearity $\xi^2 \sim 0.5$ is
acceptable~\cite{TESLATDR}. Most powerful solid state laser with
$\lambda \sim 1.05$ \MKM\ can be used upto the energies $2E_0 \sim
800$ GeV. Detailed discussion of physics, and technical problem of
photon colliders can be found elsewhere~\cite{TESLATDR,NLC,GG2000}.
Below we consider only the most critical issues: luminosity, energy, laser
system.

\subsection{Current projects of photon colliders}
Parameters of the photon colliders at TESLA~\cite{TESLATDR} (as
an example) are presented in Table~\ref{summary}, for
comparison the luminosity in \EPEM\ collisions is also given. Other
parameters, constant for all energies, are:\,\,$\lambda=1.06\;\MKM,\; N=
2\times 10^{10},\; \sigma_z=0.3\,\MM,\; f_{rep}\times n_b=
14.1\;\mbox{kHz},\; \ENX/\ENY= 2.5/0.03\times
10^{-6}\;\mbox{m$\cdot$rad}, \;\;\beta_x/\beta_y = 1.5/0.3\;\MM$.\\
\begin{table}[htb]
\caption{\label{summary} Parameters of the photon collider at TESLA} 
\vspace{0.0cm}
\begin{center}
\begin{tabular}{l c c c} \hline
$2E_0\,,$ \; GeV & 200 & 500 & 800   \\ \hline
$W_{\GG,max}$ & 122 & 390 & 670 \\
$W_{\GE,max}$ & 156 & 440 & 732 \\
$\sigma_{x/y}$ [nm] & 140/6.8 & 88/4.3 & 69/3.4  \\  
b [mm] & 2.6 & 2.1 & 2.7 \\
\LEE(geom) [$10^{34}$] & 4.8 & 12 &  19 \\  
$\LGG (z>0.8z_{m,\GG\ }) [10^{34}] $ & 0.43 & 1.1 &  1.7  \\
$\LGE (z>0.8z_{m,\GE\ }) [10^{34}] $ & 0.36 & 0.94 & 1.3 \\
$\LEE (z>0.65) [10^{34}] $ & 0.03 & 0.07 & 0.095 \\ \hline
$L_{\EPEM},\, [10^{34}]$ & 1.3 & 3.4 & 5.8  \\
\end{tabular}
\end{center} \vspace*{-0.5cm}
\end{table}  
For the same energy the \GG\ luminosity in the high energy peak of the
luminosity spectrum 
\begin{equation}
\LGG(z>0.8z_{max}) \approx (1/3) L_{\EPEM},
\end{equation}
where $z=\WGG/2E_0$. Note, that cross sections in \GG\ collisions are
typically larger then in \EPEM\ by one order of magnitude. A more
universal relation $\LGG(z>0.8z_m) \approx 0.1 L_{ee}(geom)$ 
(for $k^2 = 0.4$). Expected
\GG, \GE\ luminosity spectra at TESLA can be found
elsewhere~\cite{TEL2000,TESLATDR,TEL2002}.

The \GG\ luminosity at TESLA is limited by attainable electron beam
sizes. Having beams with smaller emittances (especially the horizontal
one) one would get a higher luminosity. In order to increase the geometric
luminosity one should decrease  the $\beta$-functions as much as
possible, down to about a bunch length. In the current scheme of the
final focus it was not possible to make $\beta_x$ below 1.5 mm due
to  chromo-geometric abberations~\cite{TESLATDR}. It is not clear
whether this is a fundamental or just a temporary technical problem.    

\subsection{Ultimate luminosity of photon colliders}
Though photons are neutral, \GG\ and \GE\ collisions are not free of
collision effects. Electrons and photons are influenced by the field
of the opposite electron beam that leads to the following
effects~\cite{TEL90}:\\[1mm]
$\bullet$ in \GG: conversion of photons into \EPEM\ pairs
(coherent pair creation); \\
$\bullet$ in \GE: coherent pair creation; beamstrahlung;
beam displacement.

Beam collision effects in \EPEM\ and \GG, \GE\ collisions are
different.  In particular, in \GG\ collisions there are no
beamstrahlung and beam instabilities which limit the horizontal beam
size in \EPEM\ collisions on the level 550 (350) nm for TESLA
(NLC/JLC). A simulation, which includes all collision effects has shown
that in \GG\ mode at TESLA one can use beams with the horizontal size
down to $\sigma_x = 10$ \NM\ (at smaller $\sigma_x$ may be problems
with the crab--crossing scheme) and influence of collision effects
will be rather small~\cite{Tfrei,TEL2000,TESLATDR}. The \GG\ luminosity
(in the high energy part) can reach $10^{35}$ \CMS.  Note that now in
TESLA project $\sigma_x\approx 500$ \NM\ in \EPEM\ collisions and
about 100 \NM\ in the \GG\ collisions.  Having electron beams with much
smaller emittances one could build a photon collider factory with
production rate of new particles by a factor of 10--50 higher than at
\EPEM\ colliders. A laser cooling of electron beams is one of
the possible methods of reducting beam emittances at photon
colliders~\cite{lascool1,lascool2}, but this method is not easy.

  Note that small rate of coherent \EPEM\ pair production at TESLA
  energies is partially explained by the beam repulsion which reduces
  the field acting on the photons. For multi-TeV energies and short
  bunches such suppression is absent and photon colliders reach their energy
  limit (with adequate luminosity) approximately at the same energies
  as \EPEM\ colliders~\cite{telsb2,telmulti,telclic}.

\subsection{Technical aspects of photon colliders}

A key element of photon colliders is a powerful laser system which is
used for the $e \to\gamma$ conversion.  Required parameters are: a few
Joules flash energy, a few picosecond duration and
10--20 kHz repetition rate.

To overcome the ``repetition rate'' problem it is quite natural to
consider a laser system where one laser bunch is used for the $e \to
\gamma$ conversion many times.  At the TESLA, the electron bunch train
contains 3000 bunches with 337 ns spacing, here  two schemes are
feasible: an optical storage ring and an external optical
cavity~\cite{TEL2000,TESLATDR,TEL2002}. With the optical cavity a required
laser power can be lower than in the case of a one-pass laser by factor of
50--100. There is no detailed scheme of such laser system  yet. 

At NLC, the electron bunch train consists of 96 bunches with 2.8 sec
spacing therefore exploiting of the optical cavity is not effective.
A current solution is a one-pass laser scheme based on the Mercury
laser developed for the fusion program. The laser produces 100--200 J
pulses which after splitting to 96 pulses can be used for $e\to\gamma$
conversion of one train~\cite{NLC,TEL2002}.

 A laser system for a photon collider
can certainly be built though it is not easy and not cheap.

\section{Advanced accelerator schemes} \label{advanced}

Conventional RF linear colliders have accelerating gradients up to 150
MeV/m, corresponding lengths about 30--40 km and attainable energies
up to 5 TeV (Sec.\ref{projects}).
On the other hands, people working on plasma and laser methods of
acceleration have obtained gradients of 100 GeV/m! Some people are
thinking already about 100 TeV and even 1 PeV linear colliders
or about 1--5 TeV LC with less than one km length. 

  Certainly, new methods of acceleration will make further progress
  and find certain applications, but it is less clear  about possibility
  of super high energy colliders based on these technologies.
  
  First of all, collision effects restrict the energy of linear colliders
  at about 10 TeV (Sec.\ref{collision}); secondly, the quality of
  electron beams should be very high; and thirdly, it is very likely
  that in considerations of very high acceleration gradients some
  important effects are just missed. Driven by my curiosity and for
  self-education  I have spent some time for random check of
  these conceptions and some remarks are presented below. Situation 
  in  this field is not bad, but  some of existing proposals are
  certainly wrong.

\subsection{Plasma acceleration}
 Laser or particle beams can excite waves in plasma with a longitudinal
electrical field~\cite{tajima0}. The accelerating gradient 
\begin{equation}
G \sim mc \omega_p \sim  10^{-4}\sqrt{n_p[\mbox{cm}^{-3}]}\left(\frac{\MEV}{\mbox{m}}
 \right).
\label{grad}
\end{equation}
Typical parameters considered: $n_p \sim 10^{15}\;\mbox{cm}^{-3}$, $G
\sim 2$ GeV/m.
\subsubsection{Multiple scattering}
Let us consider the case $n_b \gg n_p$ when all plasma electrons are
pushed out from the accelerated beam. The beams travel through ions
with density $n_p$ and experience a plasma focusing with the
$\beta$-function~\cite{whittum} $\beta \sim \sqrt{2\pi\gamma/r_e n_p} =
\sqrt{2\gamma} \lambda_p$. The r.m.s. angle due to multiple scattering 
\begin{equation}
\Delta \theta ^2 \approx \frac{8\pi Z^2 r_e^2 n dz}{\gamma^2}
\frac{d\rho}{\rho}, \;\;\;\rho_{min} \sim R_N, \;\;\rho_{max} \sim R_{D},
\end{equation}
where $R_D=(kT/4\pi n e^2)^{1/2}$ is the Debai radius. The increase of the
normalize emittance $\Delta \EN^2 = \gamma^2 r^2 \Delta \theta^2 =
\EN\ \gamma \beta \Delta \theta^2$.
After integration on the energy we get the final normalized emittance  
\begin{equation}
\EN \sim 8\pi\sqrt{2\pi} Z^2 (n_p r_e^3 \gamma_f)^{1/2} 
(mc^2/G)\;L,
\end{equation}
where $L=\ln{\rho_{max}/\rho_{\min}} \sim 20$. Substituting
$n=10^{15}\; \mbox{cm}^{-3}$, $G=2$ GeV/m, $Z=1$, $\gamma_f = 5\times
10^{6}$ ($2E_0 = 5$ TeV) we get $\EN \sim 3\times 10^{-7}$ cm.  Note
that the result does not depend on the plasma density because $G
\propto \sqrt{n_p}$ (Eq.\ref{grad}).  In present LC designs the
minimum vertical emittance $\ENY = 2\times 10^{-6}$ cm, so multiple
scattering in an {\it ideal} plasma accelerators look acceptable. It is
assumed that sections with plasma have small holes for beams since any
windows will give too large scattering angles.
\subsubsection{Synchrotron radiation}
Due to a strong focusing by ions (plasma electrons are pushed out from
the beam), beam electrons lose their energy to radiation, the radiation
power $P = (2/3) c r_e^2 \gamma^2 E_{\perp}^2$, where $ E_{\perp} =
2\pi e n_p Z r$ (as before we assume $n_b \gg n_p$), $r \sim
\sqrt{\EN\beta/\gamma}$, $\beta = \sqrt{2\pi\gamma/r_e n_p}$.  After
integration on the energy we find the difference of energies for the
particle on the axis (no radiation) and one at the r.m.s distance form
the axis
\begin{equation}
\Delta E/E \sim 25 r_e^{5/2} n_p^{3/2} Z^2 \gamma_f^{3/2}
(mc^2/G) \EN\;. 
\end{equation}
For $G=2$ GeV/m, $n_p = 10^{15} \; \mbox{cm}^{-3},\; \ENX\ \sim 10^{-4}$ cm
(emittance from damping rings or from photo-guns), $\gamma_f = 5\times
10^{6}$ ($2E_0 = 5$ TeV) we get $\Delta E/E \sim 10^{-3},$ that is
acceptable.  For several times larger energy spreads there are
chromaticity problems in final focus systems. Note, that
$G\propto \sqrt{n_p}$, therefore the energy spread is proportional to
the plasma density. In Ref.~\cite{barleta} the case of the overdense
plasma ($n_b < n_p$) was also investigated with the conclusion that it
is not suited for TeV colliders.

 So, synchrotron radiation puts a limit on a maximum plasma density
(and acceleration gradient). A 10 TeV collider with a gradient 10 times
larger that at CLIC is still possible. 
\subsubsection{Some other remarks on plasma acceleration}  
Though plasma accelerators pass the  simple criteria discussed above,
there are many other question.

At $E_0= 1$ TeV, $n_p=10^{15}$ cm$^{-3}$, \ENY=$10^{-6}$ cm the
transverse electron beam size in plasma is about $0.1$ \MKM\ ! This
means that the accelerating sections should have relative accuracy
better than $0.1 \sigma_y \sim 10^{-6}\;$ cm~!

The beam axis is determined in large extent by the drive beam. The
transverse size of the drive beam (of it's head which is focused by
external quadrupoles) is of the order of $10^{-2}$ cm. Small
fluctuations in the beam profile will lead to dilution of the
accelerated beam emittance.

To avoid radiation of the high energy accelerated beam in the field of
kickers during the injection of the drive beam, the kickers should be
very fast, in 100 GHz frequency range~\cite{rozenzweig}. Then the
required stability of the horizontal angle is about $\sim
0.1\sigma_x/L \sim 0.1\times 10^{-4}/100 \sim 10^{-7}$ rad.  If the
kick angle is $10^{-2}$ rad, the required time stability is about
$\sim \Delta x/c \times 10^{-5} \sim 10^{-16}$ sec.

In summary: a plasma acceleration is a perspective technique, it
can find certain applications, but technical feasibility of plasma
based linear colliders is not clear now. 
\subsection{Laser acceleration in vacuum} 
There are many ideas on this subject.~\footnote{In Secs 6.2, 6.3
  \CE\ denotes an electron energy and $E$ is the strength of a laser
  field.} In general, in a space with some boundaries the accelerating
gradient is proportional to the electric field, $G \propto E$, and $G
\propto E^2$ in an open space.  This is because the charged particle
extracts the energy from the field due to interference of the external
and radiated field:
\begin{equation}
\Delta \CE\ \propto \int (E + E_{rad})^2\;dV - E^2\;dV \propto E
E_{rad} \;dV,
\end{equation}
where $E_{rad} \sim const$ when a particle radiates in a given
structure  without any field  and in a free space the radiated field
is proportional to a particle acceleration: $E_{rad} \propto E$.
\vspace{-2mm}
\subsubsection{$\mathbf {G\propto E}$} 
\vspace{-2mm}
In this case, a particle is accelerated by the axial electrical field
$E_z$ of a focused laser. For a radially polarized Gaussian beam $E_z \sim
E (\lambda/\pi w_0)$, where $E$ is the transverse laser field,
$w_0$ is the radius of the focal spot.

The electron is in the accelerating phase of the wave on the length
$\sim Z_R$ (Rayleigh length). In order to get a net acceleration one
has to put some screen with a small hole to restrict the interaction
length.  The damage threshold of the optical components is a limiting
factor of the method. For the damage threshold $5$ TW/cm$^2$ the
maximum energy gain $\Delta E(\MEV) \sim 20 [P(\mbox{TW})]^{1/2}$,
that is about 50 MeV for P = 10 TW~\cite{esarey}. There is a proposal to
study this method at SLAC~\cite{E163}.

One of the other approaches uses small cavities pumped by a
laser~\cite{Mikhailichenko}. This method needs very small beam sizes
(emittances) and severe tolerances. 

There are many laser accelerating schemes of such kind under development.
\vspace{-2mm}
\subsubsection{ $\mathbf {G \propto E^2}$} 
\vspace{-2mm}
There are many fantasies on this subject. \\
{\bf 1. Light pressure}. If an electron is in the rest, a plane
electromagnetic wave pushes it with a force $F = \sigma_T \times
(E^2/4\pi)$, where $\sigma_T$ is the Thomson cross section. 
This is because laser photons scatter isotropically and  momenta of
laser photons incident onto $\sigma_T$ are transfered to the electron.
At $I= 10^{28}$ W/cm$^2$ (in Ref.\cite{tajima}), the
accelerating gradient  $d\CE/dx = 1$ TeV/cm.

Unfortunately, in a real wave, laser photons have the divergence
$\theta \sim \sqrt{\lambda/4\pi Z_R}$. Therefore, if the electron has
$ \CE/mc^2 > 1/\theta,$ then in the electron rest frame laser photons
come from the forward hemisphere and therefore the electron is
deaccelerated! For
$\lambda=1$ \MKM\ and $Z_R= 100$ \MKM, $E_{max} \sim 15$ MeV only! \\[2mm]
{\bf 2. Ponderomotive acceleration.} In a strong laser field an
electron experiences a collective force from the whole laser bunch, so
called a ponderomotive force~\cite{kibble,mora,telnov_wire},
\begin{equation}
F_i \sim \frac{mc^2}{\gamma} \frac{d\xi^2}{dx_i},\;\;\;\; \xi^2 =
\frac{e^2\overline{E^2}}{m^2c^2\omega_0^2} = \frac{2 n_{\gamma}
  r_e^2 \lambda}{\alpha}.
\end{equation}
This opens a way to transfer the energy from large body (laser beam) to
one microscopic particle (electron).  There is an idea~\cite{Smetanin}
to collide the laser pulse propagating in a rare gas (to have $v^*<c$,
large effective $\gamma^*$) with the oncoming electron bunch with a
relativistic factor $\gamma_0$, so that after elastic reflection the
electron will have $\gamma=(\gamma^*)2/\gamma^0$.  According to above
Refs, for the laser power 4.3 EW (EW=$10^{18}\;$W) $\gamma^* =
1.6\times 10^{6}$ and $\gamma_0=1400$, the energy of reflected
electrons in the laboratory system is 1 PeV $\equiv$ 1000 TeV! The
length of the collider is the laser bunch length or almost ZERO!

Unfortunately, the idea is wrong due to many reasons:\\[-6mm]
\begin{itemize}
\item The interaction length is not the bunch length but $L_{int} \sim
  l_{laser}/(1-v^*/c)  \sim l_{laser} \times (\gamma^*)^2 \sim
  10^{-2} \times 10^{12} \sim 10^{5}$ km ! 
\item  Radiation of electrons (see below), and many other ``NO''.
\end{itemize}
\subsubsection{Radiation during a ponderomotive acceleration}
During the ponderomotive acceleration electrons radiate in the
transverse laser field. This can be treated as  Compton scattering.
Radiated energy per unit length: $d\CE/dx \sim \bar{\epsilon}\;
n (1-\cos{\theta})\sigma_T.$
Substituting $\theta^2 \sim \lambda/(2\pi Z_R)$,  $\bar{\epsilon}\sim \omega_0
\gamma^2 \theta^2$,  $n=\alpha \xi^2/(2 r_e^2 \lambda)$ we get 
\begin{equation}
\frac{d\CE}{dx} \sim \frac{\xi^2 \gamma^2 r_e}{Z_R^2} mc^2.  
\end{equation}
For example: $\CE_0 = 1$ TeV, $Z_R=100\;\MKM,$ and $\xi^2=100$ (flash
energy $\sim 100$ J), $d\CE/dx = - 200$ GeV/cm.  For the mentioned 1
PeV project with $\xi^2 = 2\times 10^{6}$,  $d\CE/dx =
-10^9$ PeV/cm\,!

So,  ponderomotive acceleration can be useful for low energy
application, but not for linear colliders due to the decrease of the
force with the increase of the energy and a huge radiation.
\section{Conclusion} 
 Linear \EPEM, \EMEM, \GG, \GE\ colliders are
 ideal instrument for study of matter in the energy region
  $2E_0 \sim$ 100--1000 GeV. Three projects are almost ready for
  construction, a wise choice and political decision are needed.
  
  A linear collider is not a simple machine, 
  very high accuracies, stabilities and cleaver beam diagnostics are
  needed. Many critical elements have been tested experimentally.  

According to present understanding a
  maximum attainable energy of linear colliders with adequate
  luminosity is about $2E_0 \sim 5$ TeV. There is technology for such
  ``last'' LC, that is CLIC.
  
  Advance technologies (plasma, laser) can give higher accelerating
  gradients but their application for high energy linear colliders is
  under big question.  Further complex studies of new accelerating
  methods in this context  are needed.
\section*{Acknowledgement}
I am grateful to Pisin Chen for a great work on organization of series
of workshops on Quantum Aspects of Beam Physics, which motivated
people to look deeply in topics related to beam physics at Earth and
Cosmos and  to Atsushi Ogata for organization of the present workshop
in Hiroshima.

This work was supported in part by INTAS (00-00679).

%\newpage

\end{document}